\documentclass[journal]{IEEEtran}
\ifCLASSINFOpdf
\else
\fi
\usepackage{graphicx}
\usepackage{multirow}
\usepackage{ulem}
\usepackage{cite}
\usepackage{longtable,lscape}
\usepackage{pdflscape}
\usepackage{rotating}
\usepackage{mathptmx} 
\usepackage{times} 
\usepackage{amsmath} 
\usepackage{amssymb}  
\usepackage{multirow}
\usepackage{color}
\usepackage{xcolor} 
\usepackage{float}
\usepackage{diagbox}
\usepackage{caption}
 \usepackage{booktabs}

\newcommand{\revision}[1]{\textcolor{black!72!black}{#1}}

\newcommand{\RNum}[1]{\uppercase\expandafter{\romannumeral #1\relax}}
\begin{document}
\title{Artificial Lateral Line Based Relative State Estimation for Two Adjacent Robotic Fish}
\author{Xingwen Zheng$^{1}$, Wei Wang$^{2,3}$, Liang Li$^{1}$, and Guangming Xie$^{1,4,5}$
\thanks{$^{1}$Xingwen Zheng, Liang Li, and Guangming Xie are with the State Key Laboratory for Turbulence and Complex Systems, Intelligent Biomimetic Design Lab, College of Engineering, Peking University, Beijing, 100871, China.
        {\tt\small \{zhengxingwen, liatli, xiegming\}@pku.edu.cn.}}
\thanks{$^{2}$Wei Wang is with the SENSEable City Laboratory, Massachusetts Institute of Technology, Cambridge, MA, 02139, USA. {\tt\small wweiwang@mit.edu.}
}
\thanks{$^{3}$Wei Wang is with the Computer Science and Artificial Intelligence Lab (CSAIL), Massachusetts Institute of Technology, Cambridge, MA, 02139, USA.
}
\thanks{$^{4}$Guangming Xie is with the Peng Cheng Laboratory, 518055 Shenzhen, China.
}
\thanks{$^{5}$Guangming Xie is with the Institute of Ocean Research, Peking University, Beijing, 100871, China. Corresponding author: Guangming Xie.
}}

\maketitle
\begin{abstract}
The lateral line enables fish to efficiently sense the surrounding environment, thus assisting flow-related fish behaviours. Inspired by this phenomenon, varieties of artificial lateral line systems (ALLSs) have been developed and applied to underwater robots. \revision{This article focuses on using the pressure sensor arrays based on ALLS-measured hydrodynamic pressure variations (HPVs) for estimating the relative state between two adjacent robotic fish with leader-follower formation. The relative states include the relative oscillating frequency, amplitude, and offset of the upstream robotic fish to the downstream robotic fish, the relative vertical distance, the relative yaw angle, the relative pitch angle, and the relative roll angle between the two adjacent robotic fish. Regression model between the ALLS-measured and the mentioned relative states is investigated, and regression model-based relative state estimation is conducted. Specifically, two criteria are proposed firstly to investigate not only the sensitivity of each pressure sensor to the variations of relative state but also the insufficiency and redundancy of the pressure sensors. And thus the pressure sensors used for regression analysis are determined. Then four typical regression methods, including random forest algorithm, support vector regression, back propagation neural network, and multiple linear regression method are used for establishing regression models between the ALLS-measured HPVs and the relative states. Then regression effects of the four methods are compared and discussed. Finally, random forest-based method, which has the best regression effect, is used to estimate relative yaw angle and oscillating amplitude using the ALLS-measured HPVs and exhibits excellent estimation performance. This work contributes to local relative estimation for \revision{a group of underwater robots}, which has always been a challenge.}
\end{abstract}


\IEEEpeerreviewmaketitle

\section{Introduction}
\IEEEPARstart{L}{ateral} line system is a sensory organ system existing in most fish and aquatic amphibians. It includes superficial neuromasts on the surface of the skin and canal neuromasts in subepidermal canals. Fish can detect the surrounding flow environment using its lateral line system. It has been demonstrated that the lateral line system assists fish in varieties of behaviours like rheotaxis, schooling, obstacle avoidance, and predation \cite{Bleckmann1986Role}.

Recently, varieties of artificial lateral line systems (ALLSs) have been developed for their great potentials in improving the performance of underwater robots and surface vehicles, including remotely operated vehicle (ROV) \cite{chin2018robust}, autonomous surface vehicle (ASV) \cite{gu2019antidisturbance,wang2019path}, bio-inspired aquatic system \cite{yu2019motion}, etc. The lateral line inspired researches cover from localization of dipole source \cite{Ahrari2017Reliable,Ji2018Resolution,Abdulsadda2013Underwater,ji2019performance,Boulogne2017Performance}, identification of flow characteristics \cite{Ben2018Bio,Kanhere2016Crocodile,Chen2017Estimation}, and applications on underwater robots \cite{kruusmaa2014filose,lagor2013bio,zhang2015distributed,free2018model,asadnia2015mems,yen2018controlling,yen2018phase,zheng2017artificial,Asadnia2013Flexible,Dusek2013Development,zheng2020online}. Among the above-mentioned researches, investigating the applications of ALLS on underwater robots has been capturing \revision{an increasing attention}. Kruusmaa \emph{et al.} have investigated rheotaxis, station holding, trajectory tracking, and swimming efficiency increasing of a fish-like robot with the aid of ALLS-measured flow variations \cite{kruusmaa2014filose}. Paley \emph{et al.} have studied speed evaluation and position control of a robotic fish based on an established flow field model and flow variations measured by the ALLS \cite{lagor2013bio,zhang2015distributed,free2018model}. Tan \emph{et al.} have extensively explored ALLS based fixed and moving dipole source localisation using tank experiments and fluid dynamics \revision{simulations} \cite{Ahrari2017Reliable,Abdulsadda2013Underwater}. Guo \emph{et al.} have implemented the estimation of angular acceleration and rectilinear speed of a robotic fish using the ALLS-measured HPVs. They have also realized wall following of the robotic fish by using the ALLS-measured flow variations to estimate the relative distance and angle to the wall \cite{yen2018controlling,yen2018phase}.  Kottapalli \emph{et al.} have investigated flow-related trajectory control of a robotic stingray and realized ALLS based obstacle avoidance and navigation of an underwater vehicle \cite{asadnia2015mems,Asadnia2013Flexible,Dusek2013Development}.

The above works have demonstrated the great application potentials of ALLS in underwater robots. However, the existing works have mainly studied the application of ALLS on one individual robot, rarely investigating an underwater robot group composed of two or more individuals with ALLS. On the other hand, the flow stimuli used in the above works typically include K{\'a}rm{\'a}n vortex street (KVS) behind a cylinder or cuboid, wave generated by a vibrating sphere, and uniform flow generated by well-controlled laboratory conditions. However, these flow stimuli have the limitations to emulate the hydrodynamic characteristics of natural flow stimuli like fish tail-generated reverse KVS, which has been rarely explored.

On the basis of the above analyses, our group has mainly focused on studying how a downstream robotic fish uses its onboard pressure sensor arrays based ALLS to detect a more natural flow stimulus, that is, the reverse KVS generated by the oscillating tail of an upstream robotic fish, and then evaluates its relative states to its adjacent upstream robotic fish \cite{Wang2015Sensing,zheng2017artificial}. Specifically, we have conducted massive flume experiments using two adjacent robotic fish with leader-follower formation in a flume. The KVS generated by the upstream robotic fish causes the hydrodynamic pressure variations (HPVs) in the flow field. Using ALLS, the downstream robotic fish can detect the pressure variations, thus detecting the reverse KVS indirectly. By extracting meaningful information from the ALLS readings, \revision{the qualitative relationships between the HPVs and the relative states including the relative oscillating frequency $f$, oscillating amplitude $A$, and oscillating offset $\phi$ of the upstream robotic fish to the downstream robotic fish, the relative vertical distance $\revision{d}$, the relative yaw angle $\alpha$, the relative pitch angle $\beta$, and the relative roll angle $\gamma$ between the two adjacent robotic fish have been obtained \cite{zheng2017artificial}.}

\revision{Previous studies \cite{zheng2017artificial,Wang2015Sensing} have mainly focused on the experiments, including experimental conditions, the experimental platform, the method of measuring the ALLS data, the visualisation of the ALLS data, and the explanation of data regularities. However, no models between the ALLS data and the relative states have been built. Thus the results can not be directly used for realising the relative state estimation of two adjacent robotic fish. While this article has mainly focused on the applications of the ALLS-measured flow variations in estimating the relative states, which is essential to flow-aided formation control of underwater robot group in the future. Specifically, we have focused on establishing a hydrodynamic model which refers to regression model linking the ALLS-measured HPVs of downstream robotic fish to the relative states.}

\revision{However, only a few works \cite{franosch2009wake,ren2012model,akanyeti2011information,zheng2020online} have investigated \revision{modeling for hydrodynamic variations} and it is always a challenge to precisely characterise the flow variations caused by the vortices. Potential flow theory has provided a promising approach for quantitatively describing the flow variations existing in fish tail-generated vortex wake \cite{franosch2009wake} and surrounding fish body \cite{zheng2020online}, thus establishing the above-mentioned regression model. However, the existing works have only focused on one individual fish or fish robot \cite{franosch2009wake,zheng2020online}. The hydrodynamic variations modelings for two or more fish robots are quite different because it is extremely difficult to establish a model as reference, basing on flow dynamics theories \cite{zheng2019data}. In this case, this article attempts to investigate a regression model between the ALLS data and the relative states using intelligent algorithms.}

\revision{Before conducting the regression modeling, two criteria are proposed firstly for investigating the sensitivity of each pressure sensor to the variations of relative state, then the pressure sensors are sorted according to their sensitivities. Basing on the order, the insufficiency and redundancy of the pressure sensors are analysed in detail. Then the reasonable number of pressure sensors used for regression analysis is determined. To \revision{the best of} our knowledge, no works have investigated the sensitivity of sensors in ALLS, and only a few works have analysed insufficiency and redundancy for sensor arrays based ALLS \cite{devries2013observability}. Such an investigation is helpful for critical for final robotic applications, especially for obtaining the best efficiency of the regression models using the least sensors, and thus decreasing the time of data processing when applying the regression model in online estimation.}

\revision{Besides, because the regression model between the relative state and the flow variations around the robotic fish is unknown. We can not determine which is the best method for investigating the above regression models. So four \revision{typical regression methods which have shown great performance in regression analysis of variables are used for establishing the regression model}. The methods include random forest (RF) algorithm, support vector regression (SVR), back propagation neural network (BPNN), and multiple linear regression method (REG). And by comparing the regression effects of using the four methods in detail, the RF method has been determined as the best method. Finally, the estimations for two relative states, relative yaw angle and the oscillating amplitude of the upstream robotic fish, have been conducted to verify the effectiveness of the RF method.}

\revision{The contributions of this article can be concluded as follows:}

1) This article guides the application of ALLS in local perception. Particularly, this article investigates how to obtain the relative states by reversely solving the established regression models using the ALL-measured HPVs. This article demonstrates the effectiveness of ALLS in not only close-range perception but also relative state estimation for two or more individuals in an underwater robot group, which has always been a challenge. The work also has the possible extension to underwater robot group control.

2) Quantitatively investigating multiple relative states using ALLS-measured data measured in a large scope of experimental parameter space based on various intelligent algorithms. To \revision{the best of} our knowledge, few lateral line inspired researches conducting with underwater robots have investigated various states of the robot. And the experiments have mainly conducted with a fairly limited experimental parameter space.

3) \revision{Analysing the sensitivity of each pressure sensor to the variations of relative state and the insufficiency or redundancy of the number of the sensors used when extracting information from the ALLS-measured data, such an analysis has been rarely conducted before \cite{devries2013observability}.}

The remainder of this article is organised as follows. Section II introduces the robotic fish with the ALLS. Section III describes the four methods, the criteria for investigating insufficiency or redundancy of the pressure sensors, the importance measurement of the HPVs measured by each pressure sensor, and the evaluation of the regression model. Section III shows the pretreatment results of the sample data and the regression results using the four methods. Section IV discusses the work. Section V concludes this article with an outline of future work.
\section{The Robotic Fish with an Artificial Lateral Line System}
\begin{figure}[!h]
\centering
\includegraphics[width=0.8\linewidth]{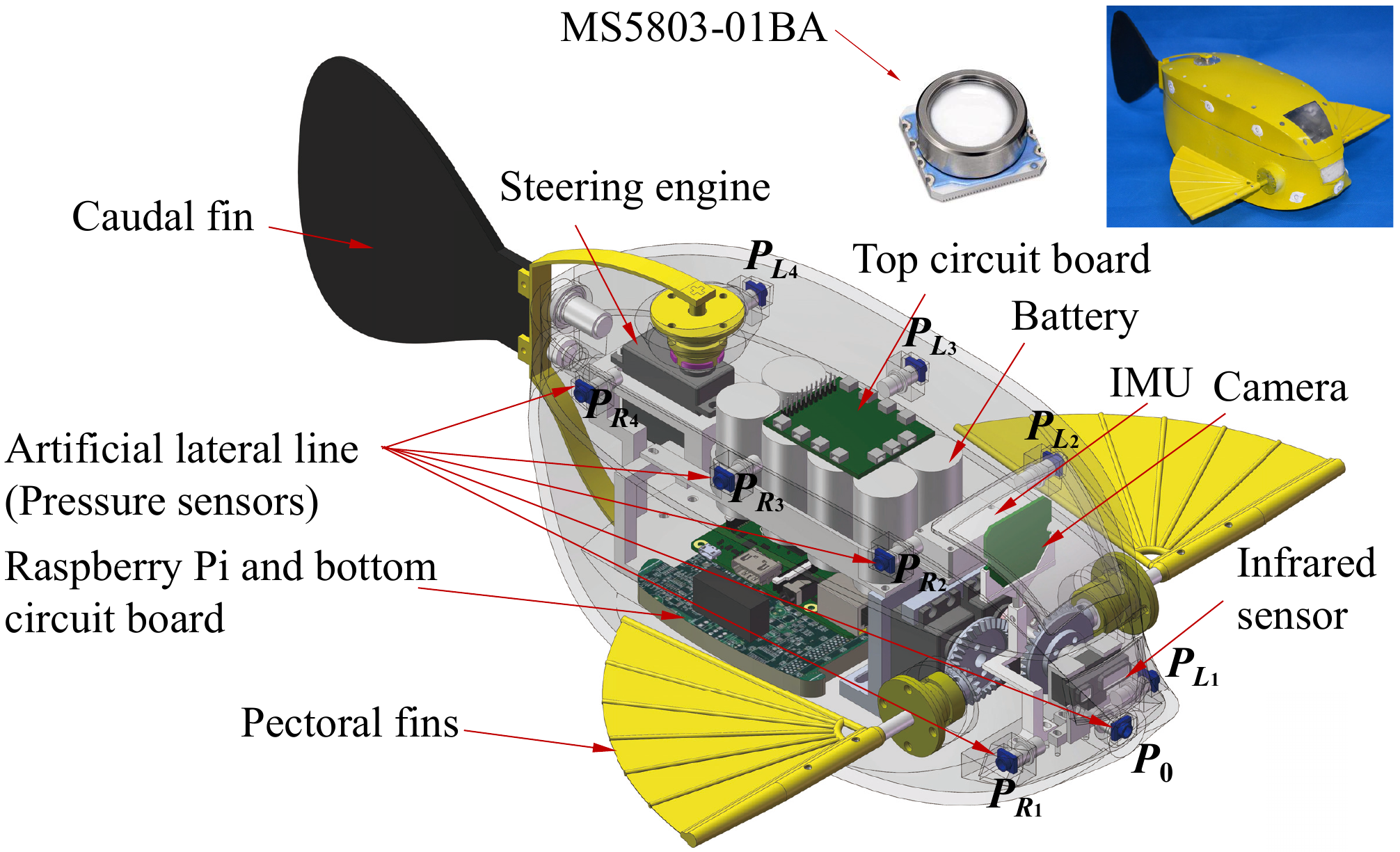}
\caption{Boxfish-like robot with an ALLS. The pressure sensors used for forming the ALLS are named $P_{L_m}$ and $P_{R_m}$ $(m=1,2,3,4)$ (Reused from \cite{zheng2017artificial} with permission of IOP Publishing).}
\label{The prototype of boxfish-like robot}
\end{figure}
\begin{figure}[!h]
\centering
\includegraphics[width=0.8\columnwidth]{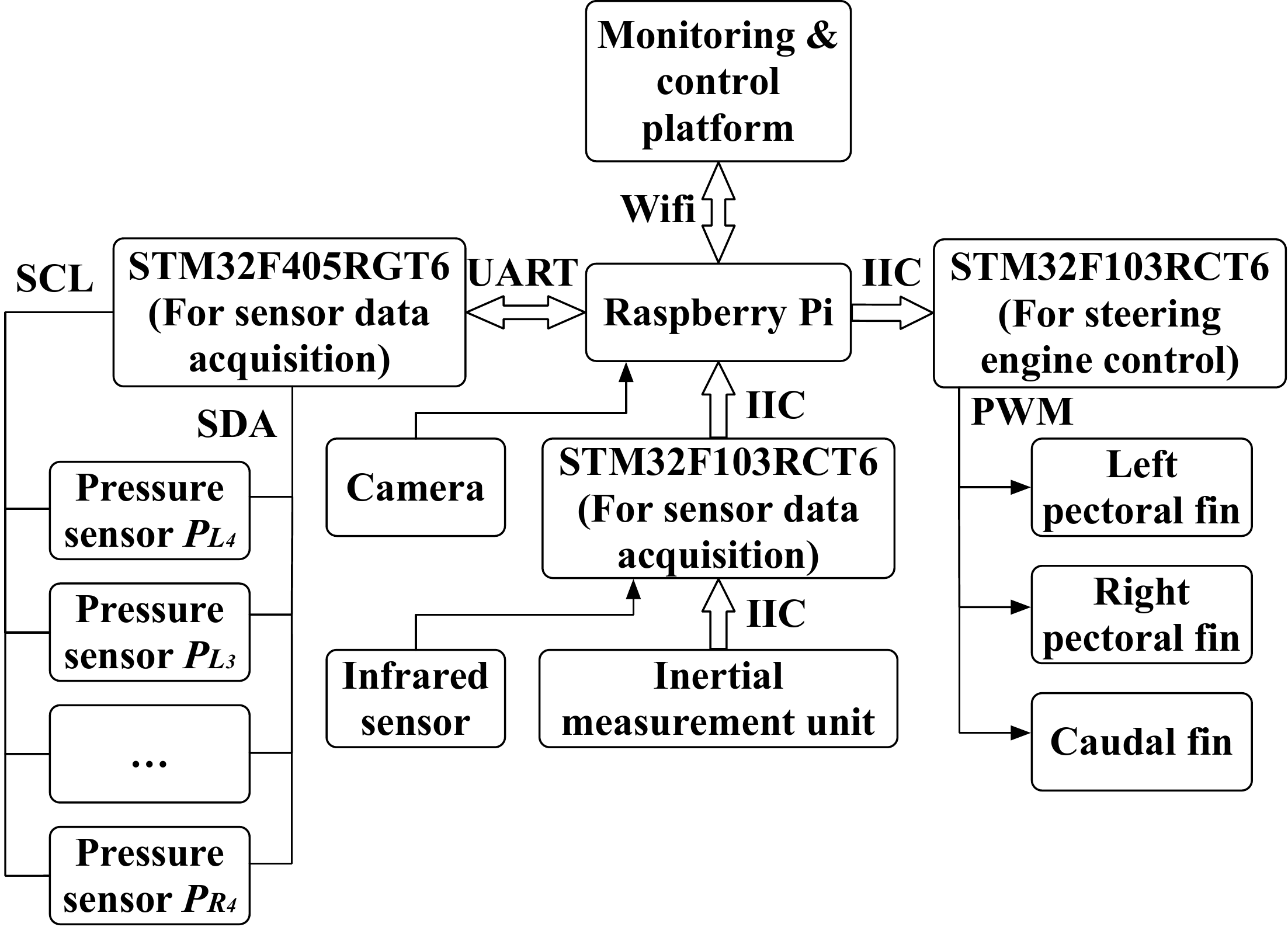}
\caption{The diagrammatic sketch of the electrical system of the boxfish-like robot.}
\label{he diagrammatic sketch of electrical system of the robotic fish}
\end{figure}
\begin{figure}[!h]
\centering
\includegraphics[width=0.8\columnwidth]{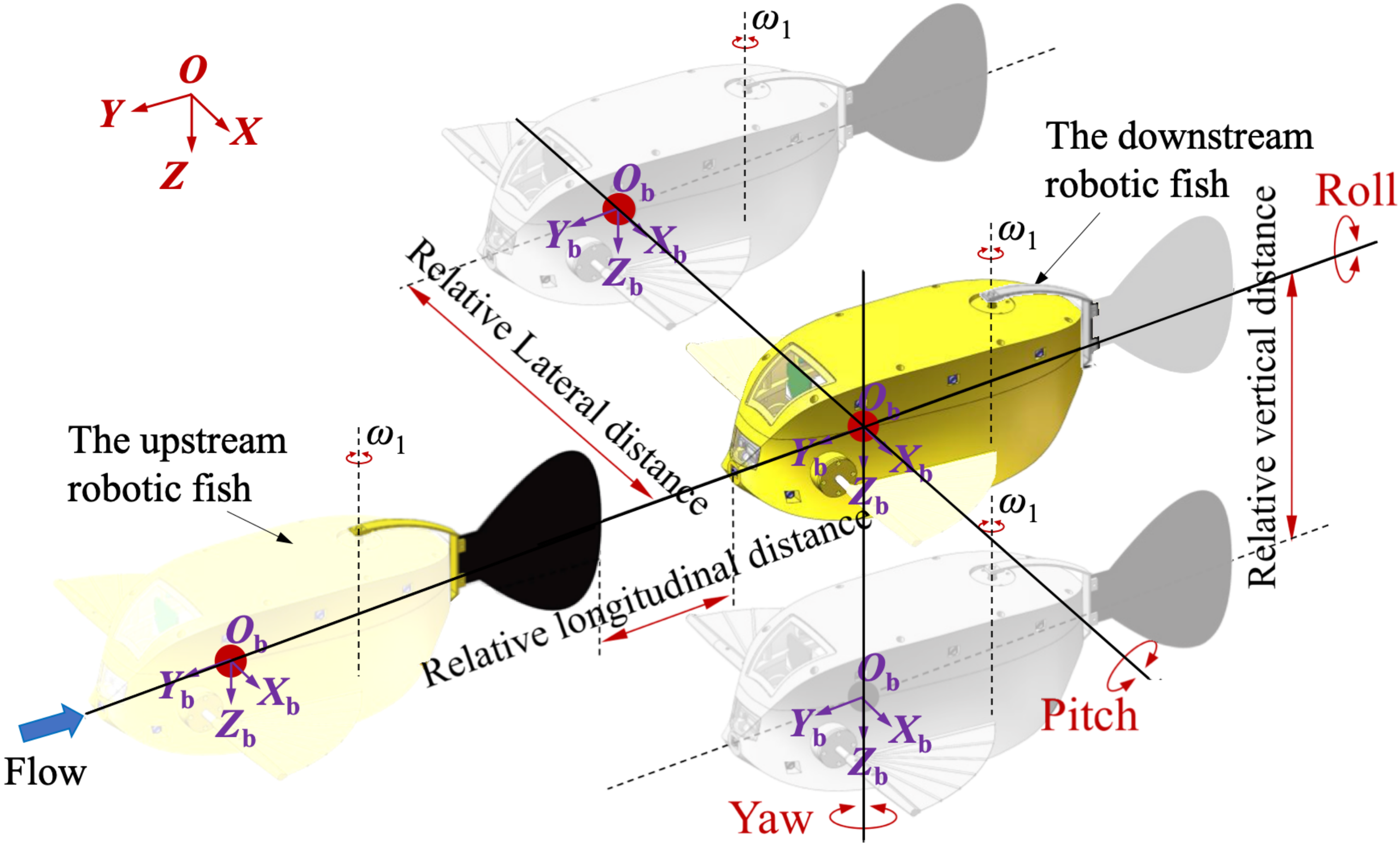}\\
(a)\\
\includegraphics[width=0.8\columnwidth]{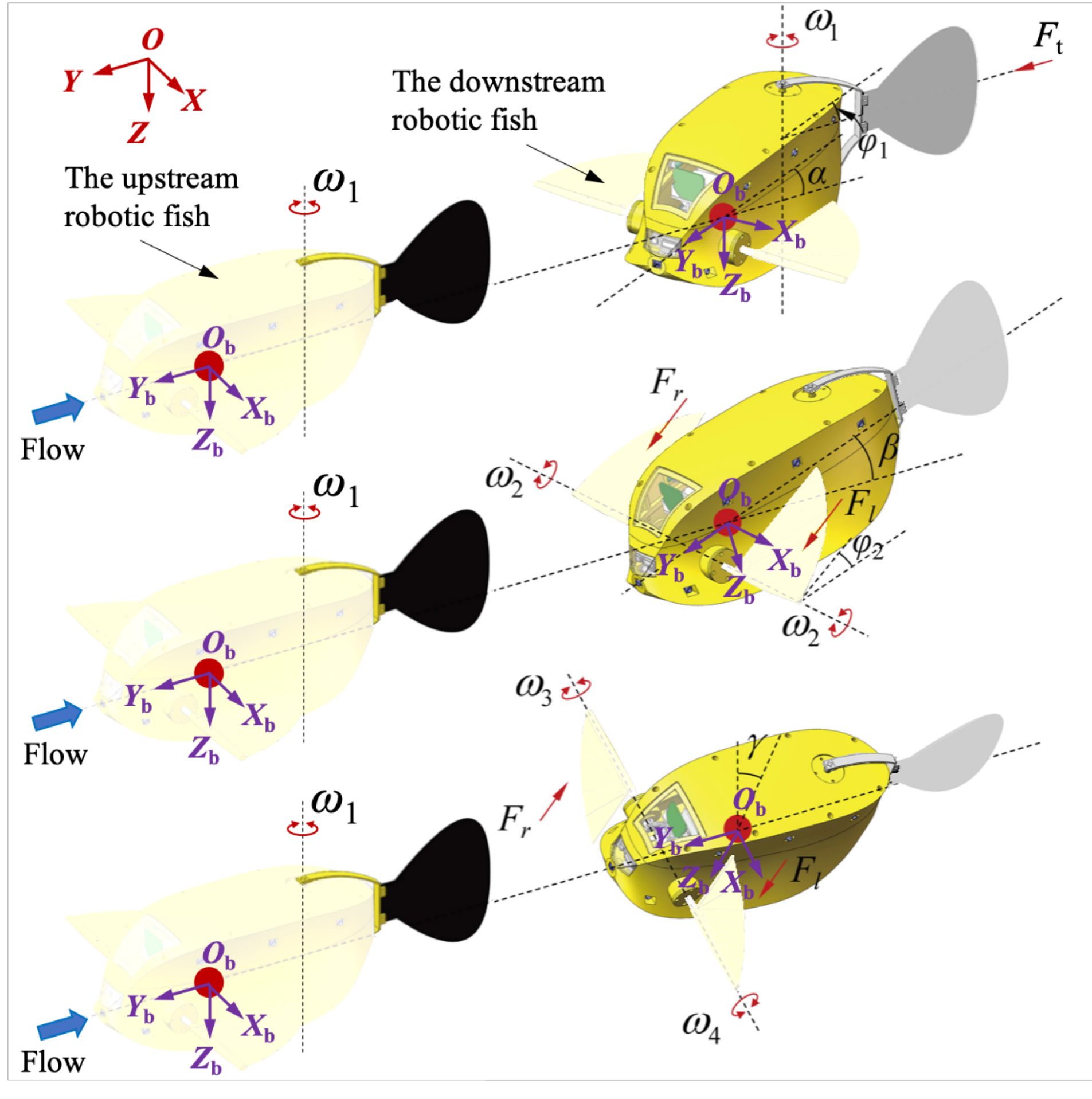}\\
(b)
\caption{Relative positions and attitudes between two adjacent robotic fish. (a) Relative positions between two adjacent robotic fish and definition of the attitude motions. The red point indicates the center of mass of the robotic fish. (b) Relative attitudes between two adjacent robotic fish. The relative attitudes include relative yaw angle $\alpha$, relative pitch angle $\beta$, and relative roll angle $\gamma$. $\omega_{1}$, $\omega_{2}$, $\omega_{3}$, and $\omega_{4}$ indicate the angular velocities of the fins. $\varphi_{1}$ and $\varphi_{2}$ indicate the offsets of the fins. $OXYZ$ and $O_bX_bY_bZ_b$ indicate the global inertial coordinate system and the fish body-fixed coordinate system, respectively. $F_t$, $F_l$, and $F_r$ indicate the propulsive force generated by the oscillating caudal fin, the left pectoral fin, and the right pectoral fin, respectively.}
\label{Relative positions and attitudes between two robotic fishes}
\end{figure}

Figure~\ref{The prototype of boxfish-like robot} shows the configurations of the robotic fish used in the experiments. It is a boxfish-inspired robotic fish whose size (Length$\times$Width$\times$Height) is about 40cm$\times$14.1cm$\times$13.2cm. It consists of a sealed shell, a pair of pectoral fins, and a caudal fin. The electrical system, which includes a credit card-sized micro-computer called Raspberry Pi serving as the main processor, rechargeable batteries, steering engines, circuit boards, and multiple onboard sensors, is wrapped in the sealed shell. As shown in Figure~\ref{he diagrammatic sketch of electrical system of the robotic fish}, the onboard sensors include an inertial measurement unit (IMU) serving for attitudes monitoring, a camera serving for environment capturing, an infrared sensor serving for obstacle avoidance, and a nine pressure sensors (MS5803-01BA, TE Connectivity Ltd) based ALLS serving for external HPVs measurement. More about the ALLS can be found in \cite{zheng2017artificial}. Multiple 32-bit micro controllers (STM32F405 and STM32F103) on the circuit boards serve the functions of sensor data acquisition and steering engine control. Through controlling the steering engines connected with the paired pectoral fins and the caudal fin with given oscillating amplitude, frequency, and offset parameters, the fins are able to generate propulsive forces which actuate the robotic fish for realizing multiple swimming patterns including rectilinear motion, turning motion, gliding motion, spiral motion, yawing motion, rolling motion, pitching motion, etc. Such multiple swimming patterns enable two adjacent robotic fish to form multiple relative positions and attitudes, as shown in Figure~\ref{Relative positions and attitudes between two robotic fishes}.
\section{Methods}
\subsection{\revision{Acquisition of Experimental Data}}
\revision{As shown in Figure~\ref{Relative positions and attitudes between two robotic fishes}, in the experiments, we have replaced the upstream robotic fish with an individual caudal fin. And the pectoral fins and caudal fin of the downstream robotic fish have been removed. The two objects are located in a flume with a specific velocity of 0.175 m/s, with given relative states. The positions and attitudes of the two objects are controlled by controllable steering engines linked with them. Thus the relative states can be set, and there are 7 relative states including $\revision{d}$, $f$, $A$, $\phi$, $\alpha$, $\beta$, and $\gamma$ in total. For each relative state, we investigate $p$ experimental parameters, and $p$=7, 16, 6, 13, 19, 9, and 11 for $\revision{d}$, $A$, $f$, $\phi$, $\alpha$, $\beta$, and $\gamma$, respectively. Details of the $p$ experimental parameters mentioned above are defined in Table~\ref{Details_of_the_$p$_experimental_parameters}. For each experimental parameter, the HPVs recording is repeated five times. As we have mentioned in detail, this article mainly focuses on investigating the sensitivity of each pressure sensor, the insufficiency and redundancy of the pressure sensors, regression modeling, and model based relative state estimation using the measure experimental data, to avoid the repetition, \revision{more details about such a simplification, the control of relative states between the two adjacent objects, the experimental conditions, the experimental platform, and the method of measuring data can be found in \cite{zheng2017artificial}}. The experimental scenes can be found in supplementary video. In the following part, we will use the ALLS-measured HPVs to estimate the above-mentioned relative states using the RF algorithm, BPNN, SVR method, and REG method.}
\subsection{Pretreatment of the Data}
\revision{Because of the hardware deficiency of the pressure sensors and the background noise in the environment, \revision{there exist significant fluctuations of the HPVs measured by the pressure sensors}. In this case, we have smoothed the raw HPVs data using a Gaussian smoothing window function. Fig. S1 in \revision{Supplemental Materials} shows the raw HPVs and the smoothed HPVs data. In each recording, 250 samples of the HPVs are selected for forming the original sample set $O$ used for regression model analysis. \revision{Taking} the regression analysis for $\revision{d}$ for example, there are seven experimental parameters of $\revision{d}$, varying from -45 mm to 45 mm, with an interval of 15 mm. The final samples in the original sample set $O$ for $\revision{d}$ have a size of $7\times5\times250=8750$.}
\begin{table}[]
\centering
\caption{\revision{Details of the $p$ experimental parameters.}}
\label{Details_of_the_$p$_experimental_parameters}
\revision{\begin{tabular}{@{}lll@{}}
\toprule
Items & $p$  & Parameters \\ \midrule
$\revision{d}$ (mm)       & 7  &  -45, -30, -15, 0, 15, 30, 45          \\
$A$ (degree)      & 16 &  \{0, 2, 4, ..., 26, 28, 30\}          \\
$f$ (Hz)     & 6  & 0.5, 0.6, 0.7, 0.8, 0.9, 1.0           \\
$\phi$ (degree)     & 13 & \{30, -25,-20,  ...,20,  25, 30\}           \\
$\alpha$ (degree)      & 19 & \{-90, -80, -70, ..., 70, 80, 90\}           \\
$\beta$ (degree)     & 9  & \{-20, -15, -10, ..., 10, 15, 20\}           \\
$\gamma$ (degree)      & 11 & \{-50, -40, -30, ..., 30, 40, 50\}           \\ \bottomrule
\end{tabular}}
\end{table}

\subsection{Random Forest for Regression Task}
Random forest, proposed by Breiman, L., is a decision-tree-based ensemble learning algorithm which can be used for classification and regression \cite{breiman2001random}. In this article, random forest is used for regression. A random forest can be defined as $R=\left \{ T_1\left ( X \right ), \dots, T_N\left ( X \right )  \right \}$, consisting of $N$ decision trees which are defined as $T_i (i=1, \dots, N)$. $X=\left \{ X(1), \dots, X(p)  \right \}$ is a $p$-dimensional feature vector. In a regression task, by importing $X$ into the random forest $R$, $N$ estimated values $\hat{Y}_i (i=1, \dots, N)$ of $Y$ are obtained, where $\hat{Y}_i$ indicates the estimated value of $Y$ obtained by the tree $T_i$. $\hat{Y}=\frac{\hat{Y}_1+\cdots+\hat{Y}_N}{N}$ is the estimated result of $Y$ using the random forest method. By importing a sample data set $D=\{ (X_1, Y_1), \dots, (X_n, Y_n)\}$ for training the random forest model, a model which links $X_j (j=1, \dots, n)$ to $Y_j$ can be obtained \cite{Liu2012New}. In this article, each sample in the original sample set consists of the HPVs measured by the nine pressure sensors of the ALLS and value of the relative state. Each sample can be defined as $\{ S, P_0, P_{L_1}, P_{L_2}, P_{L_3}, P_{L_4}, P_{R_1}, P_{R_2}, P_{R_3}, P_{R_4}\}$, where $S$ indicates value of the relative state while other variables indicate the ALLS-measured HPVs. In this article, for each sample, the feature vector $X$ includes nine HPVs measured by the ALLS, and $X=\{P_0, P_{L_1}, P_{L_2}, P_{L_3}, P_{L_4}, P_{R_1}, P_{R_2}, P_{R_3}, P_{R_4}\}$. While $Y$ is the relative state $S$.

\revision{Fig. S2  in Supplemental Materials} shows the flow of the random forest for the regression task. It includes four steps described as follows \cite{breiman2001random,Liu2012New}.

$\textbf{Step 1:}$ Obtain a bootstrap sample set of size $n$ by randomly sampling with replacement from the original sample set $O$ of size $n$, as a training set $D$ of the established random forest. The data samples which \revision{have not been sampled} into the bootstrap sample set are called out-of-bag (OOB) samples and form a validation set $V$.

$\textbf{Step 2:}$ Randomly select $m$ variables from $M$ variables of the original sample set as branch variables at each node of each regression tree. Then determine the optimum splitting attributes according to the goodness of split criterion. In general, $m=M/3\pm 1$. If $M< 3$, $m$ is set as 1. In this article, the variables of the original sample indicate nine ALLS-measured HPVs, so $M=1\sim 9$.

$\textbf{Step 3:}$ Repeat Step 1 and 2 for $N$ times, for obtaining $N$ training sets and $N$ validation sets. Then respectively establish an initial random forest consisting of $N$ regression trees $T_i (i=1, \dots, N)$ for the above training sets and validation sets.
\revision{Fig. S3  in Supplemental Materials} shows the mean of squared residuals with $N$ for $\revision{d}$, $f$, $A$, $\phi$, $\alpha$, $\beta$, and $\gamma$. It can be seen that the mean of squared residuals gradually decreases with the increasing $N$, and it converges to the minimum when $N$ exceeds 400. In this article, $N$ is set as 500.

$\textbf{Step 4:}$ Calculate mean absolute error ($MAE$) and coefficient of determination ($R^2$) of the estimated results using the original sample set $O$.

\subsection{Back propagation neural network (BPNN)}
The artificial neural network has been widely used for modeling the relationship between the input signal and output signal. It has exhibited excellent performance in tasks of data classification, data speculation, and pattern identification \cite{hsu1995artificial}. BPNN is one of the neural networks which has the widest application. The information in a BPNN propagates forward while the error propagates backward. The basic theory of BPNN is using gradient descent methods to make the mean square error between the actual output and the expected output least. For a BPNN, it consists of one input layer, one or more hidden layers, and one output layer. In each layer, there are several nodes \cite{hecht1992theory}. In this article, a three-layer BPNN has been respectively structured for each experiment, as shown in Fig. S4 of Supplemental Materials.
The BPNN has one input layer of which the number of nodes $p=9$ and the nine HPVs in each sample are used as the node data. Besides, there is one output layer of which the number of nodes $q=1$ and the relative state in each sample is used as the node data. Besides, there is one hidden layer of which the number of nodes $m$ is determined as follows.
\begin{figure}[!h]
\centering
\includegraphics[width=0.8\linewidth]{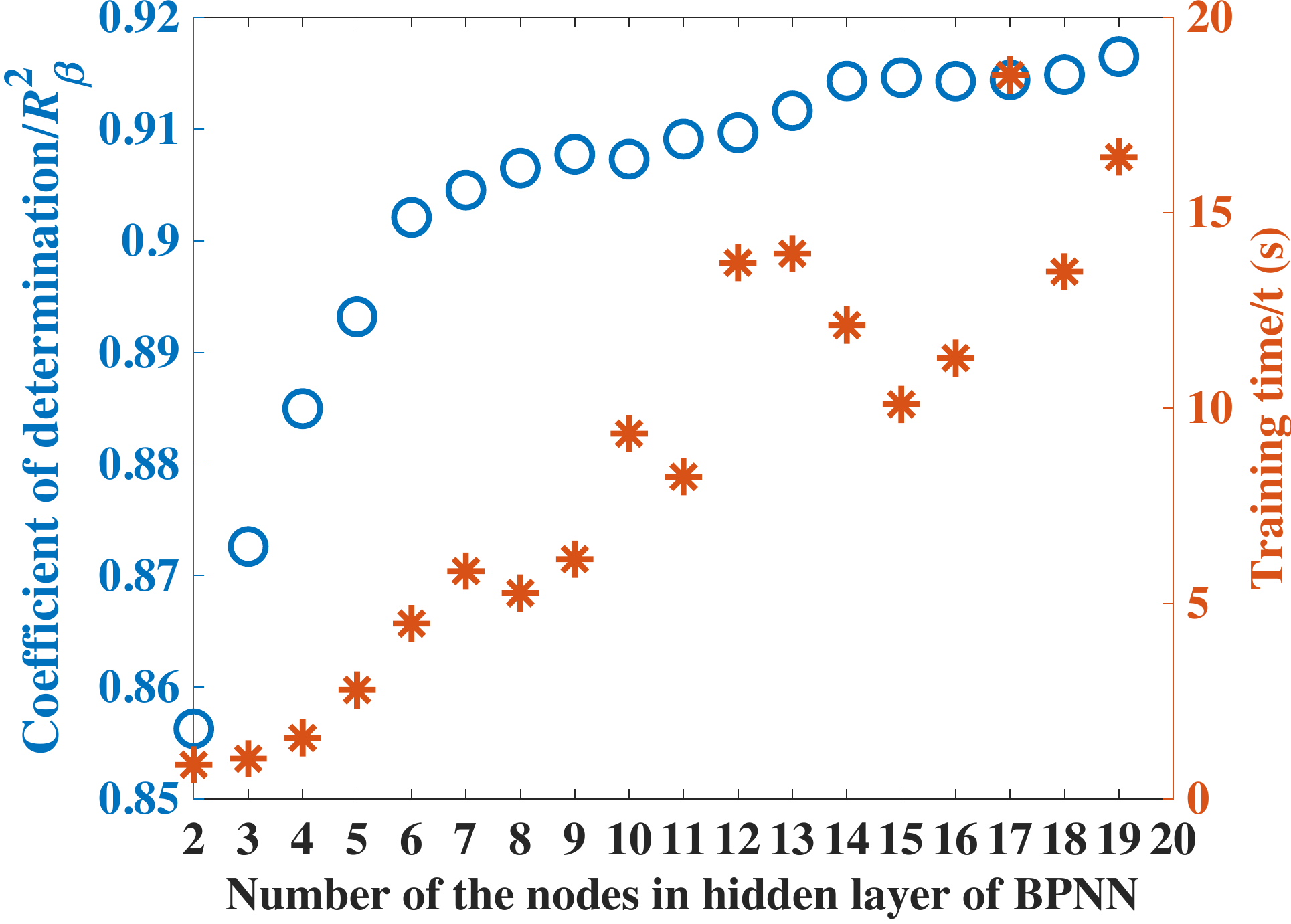}
\caption{\revision{Training time (in $*$) and coefficient of determination $R^2$ (in o) with respect to number of the nodes in the hidden layer of BPNN for the estimation of $\beta$. The number of the variables used ($M$) equals 9. The iterations of BPNN equals 1000.}}
\label{Training time bp}
\end{figure}
\begin{figure}[!h]
\centering
\includegraphics[width=0.8\linewidth]{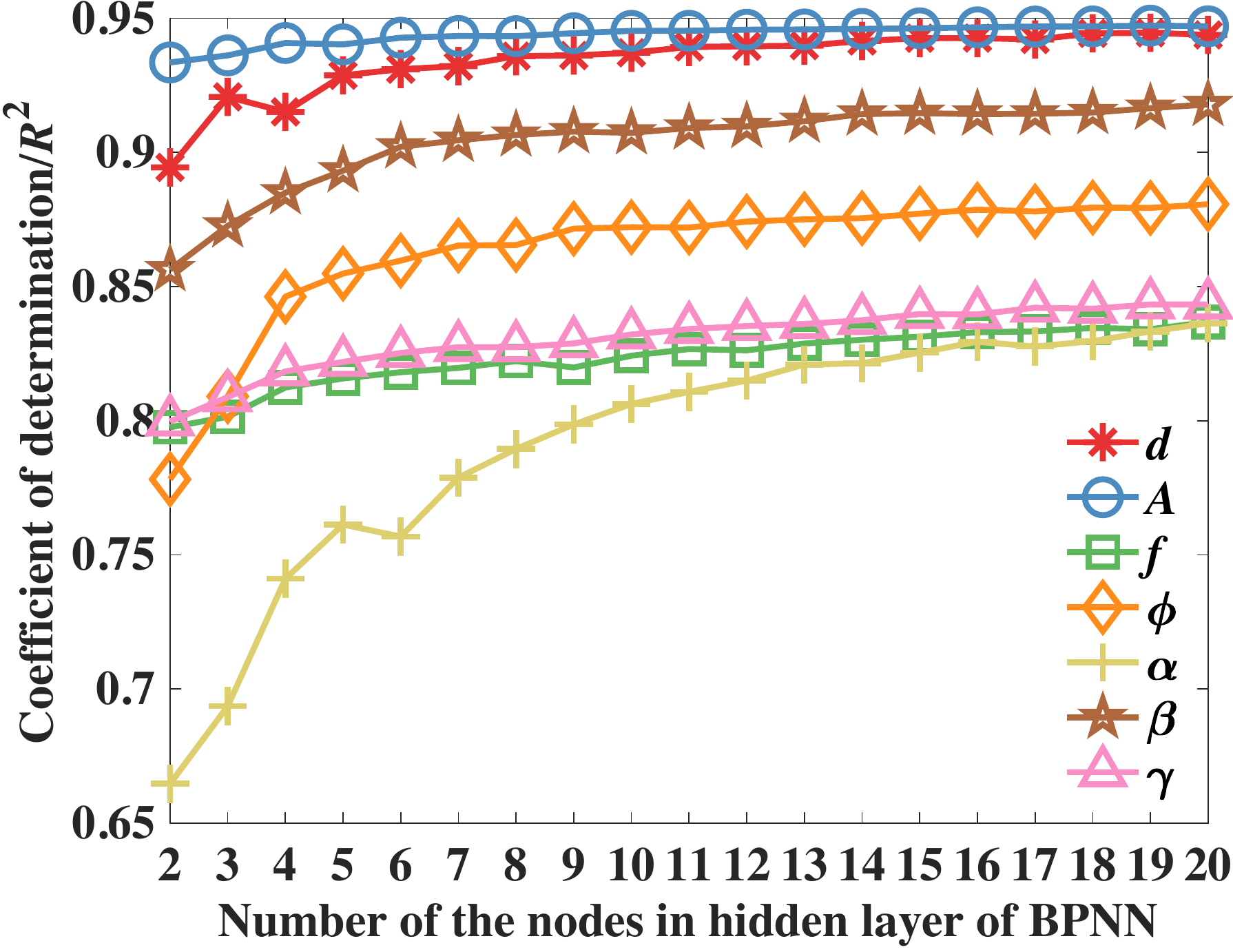}
\caption{\revision{Coefficient of determination $R^2$ with $m$ for all the experiments. Number of the variables used $M$ equals 9.}}
\label{bpnn_r2}
\end{figure}
\begin{figure}[!h]
\centering
\includegraphics[width=0.8\linewidth]{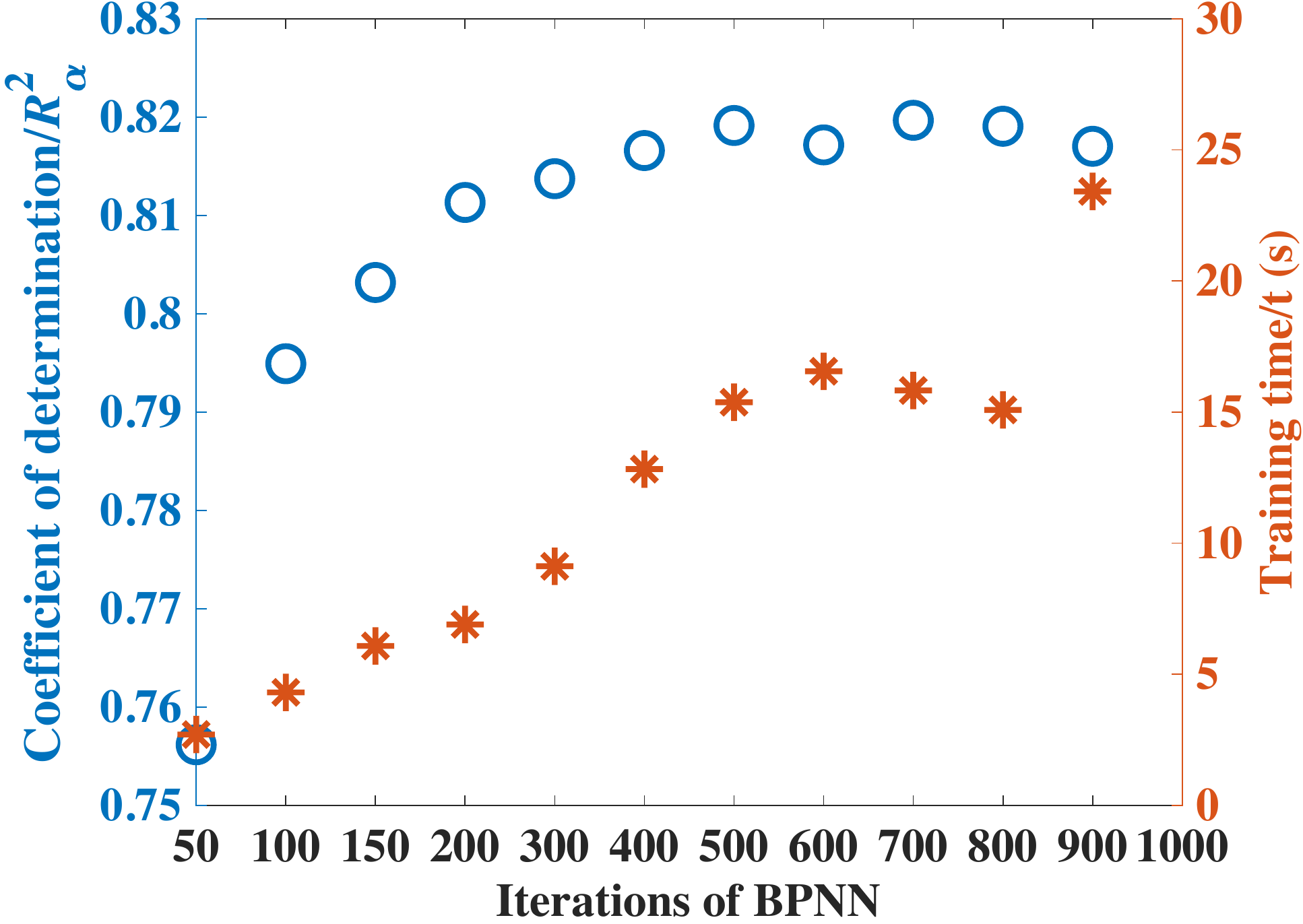}
\caption{\revision{Training time (in $*$) and coefficient of determination $R^2$ (in o) with respect to iterations of BPNN for the estimation of $\alpha$. The number of the variables used ($M$) equals 9.}}
\label{iterations}
\end{figure}
\begin{figure}[!h]
\centering
\includegraphics[width=0.8\linewidth]{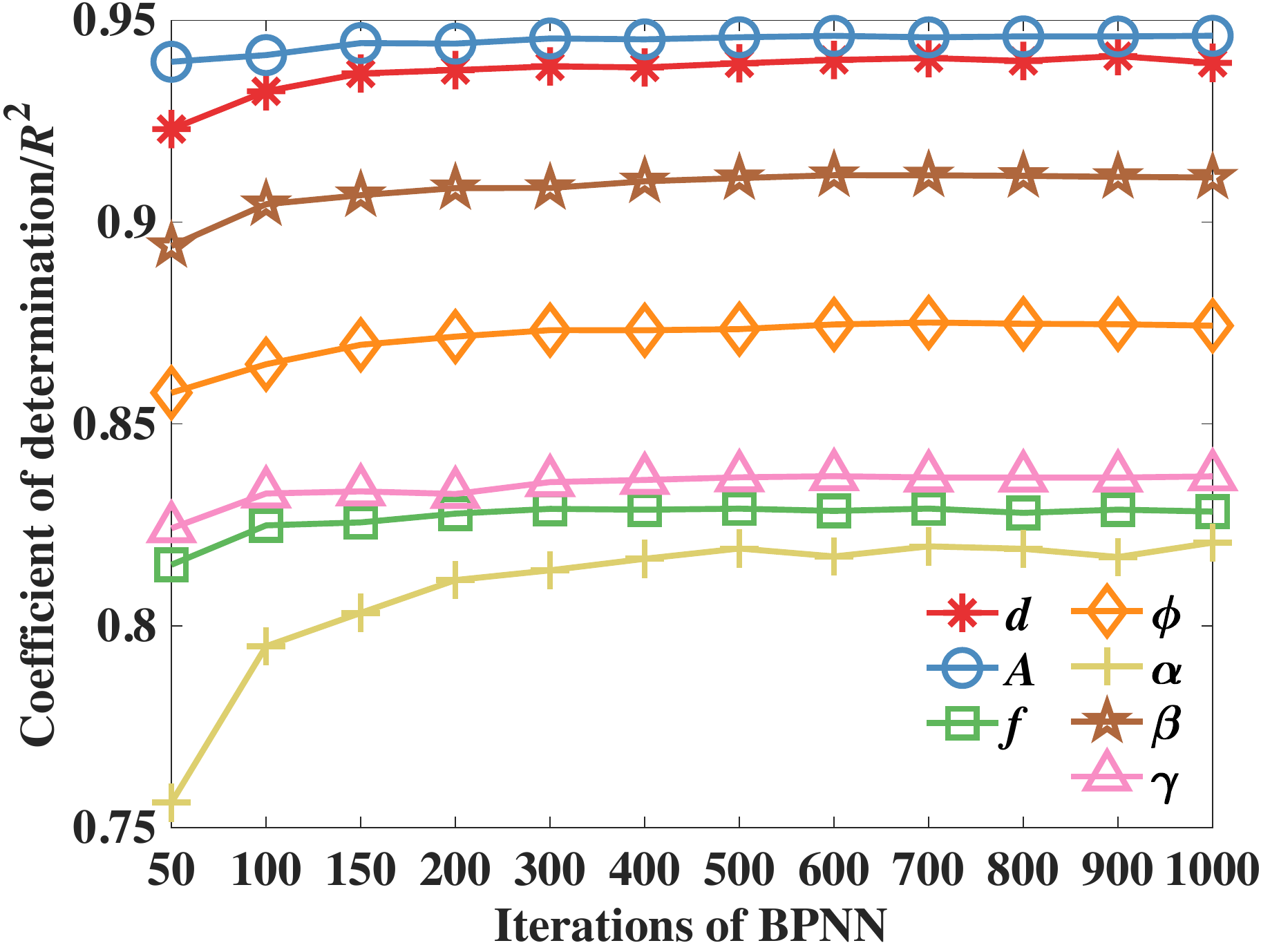}
\caption{\revision{Coefficient of determination $R^2$ with iterations of BPNN for all the experiments. The number of the variables used ($M$) equals 9.}}
\label{iterations_all}
\end{figure}

Figure~\ref{Training time bp} shows the training time and coefficient of determination $R^2$ with respect to the number of the nodes $m$ in the hidden layer of BPNN for the estimation of $\beta$. It can be seen that the training time increases with $m$, while $R^2$ varies little when the number of nodes exceeds 6. In this case, we have determined $m$ as 6 for the estimation of $\beta$. Similarly, $m$ has been determined as 11, 10, 9, 6, 13, 6, and 10 for the estimation of $\revision{d}$, $A$, $f$, $\phi$, $\alpha$, $\beta$, and $\gamma$, respectively, according to Figure~\ref{bpnn_r2}.

As shown in Figure~\ref{iterations}, the training time increases with iterations of BPNN, while $R^2$ varies little when iterations exceeds 400. In this case, iterations of BPNN for $\alpha$ is determined as 400. Figure~\ref{iterations_all} shows the coefficient of determination $R^2$ with iterations of BPNN for $\revision{d}$, $A$, $f$, $\phi$, $\alpha$, $\beta$, and $\gamma$. Similarly, iterations of BPNN for $\revision{d}$, $A$, $f$, $\phi$, $\alpha$, $\beta$, and $\gamma$ are determined as 150, 250, 150, 200, 400, 150, and 300, respectively.
\subsection{Support Vector Regression (SVR) and Multivariable Linear Regression (REG)}
\subsubsection{Support vector regression (SVR)}
SVR is used to describe regression using support vector methods. It is developed from support vector machine (SVM) by introducing an alternative loss function \cite{gunn1998support}. In this article, SVR with eps-regression and radial basis function as the kernel is used to investigate the regression relationship between the relative state and the ALLS-measured HPVs.
\subsubsection{Multivariable Linear Regression (REG)}
REG refers to investigating a linear regressive relationship between a dependent variable and two or more independent variables. In this article, we establish a model for linking the ALLS-measured HPVs to the relative states, as follows:
\begin{eqnarray}
Y=a_0+\sum a_kX(k)+\varepsilon
\label{Multivariable linear regression}
\end{eqnarray}
where $Y$ is the relative state, $a_0$ is the intercept, $a_k$ is the regression coefficient corresponding to the variable $X(k), k=1,2, \cdots, 9$, $X(k)$ indicates the HPVs measured by the nine pressure sensors, and $\varepsilon$ is the residual error.
F-test is conducted for verifying the rationality of the model. To be specific, the model has the rationality to explain the linear relationship between the dependent variable and the independent variables only when all of the regression coefficients don't equal to zero at the same time.
\subsection{Insufficiency and Redundancy of the Pressure Sensors}
In order to study the insufficiency and redundancy of the pressure sensors used for regression analysis, we have first proposed two criteria for measuring the sensitivity of the HPVs measured by the pressure sensors to the variations of relative states. The determination of the two criteria is described as follows.

$\textbf{Step 1:}$ Calculating the variations of the HPVs ($\Delta HPV$) with the variations of the experimental parameters ($\Delta E$) for each pressure sensor. The above experimental parameters include $\revision{d}$, $A$, $f$, $\phi$, $\alpha$, $\beta$, and $\gamma$.
\begin{eqnarray}
\Delta \overline{HPV_{i}(k)}=\overline{HPV_{i+1}(k)}-\overline{HPV_{i}(k)}
\label{Delta HPV1}
\end{eqnarray}
where $i=1, 2, \cdots, p-1$. $p$ is the number of the experimental parameters, and $p$=7, 16, 6, 13, 19, 9, and 11 for $\revision{d}$, $A$, $f$, $\phi$, $\alpha$, $\beta$, and $\gamma$, respectively. $\overline{HPV_{i}(k)}$ indicates the mean HPVs of 500 samples in the $i$-th experimental parameter for $X(k)$.

$\textbf{Step 2:}$ Calculating the error $\Delta \overline{HPV_{i}(k)}$ between the maximum and the minimum of $\overline{HPV_{i}(k)}$.
\begin{eqnarray}
mm\overline{HPV_{i}(k)}=max\overline{HPV_{i}(k)}-min\overline{HPV_{i}(k)}
\label{Delta HPV2}
\end{eqnarray}

$\textbf{Step 3:}$ Nondimensionalize the $\Delta \overline{HPV_{i}(k)}$.
\begin{eqnarray}
{\Delta \overline{HPV_{i}(k)}}'=\frac{\Delta \overline{HPV_{i}(k)}}{mm\overline{HPV_{i}(k)}}
\label{Delta HPV3}
\end{eqnarray}

$\textbf{Step 4:}$ Calculate the mean of ${\Delta \overline{HPV_{i}(k)}}'$ and the mean of ${\Delta \overline{HPV_{i}(k)}}$.
\begin{eqnarray}
C_{k_1}=\frac{\sum_{i=1}^{p-1}{\Delta \overline{HPV_{i}(k)}}'}{p-1}
\label{Delta HPV4}
\end{eqnarray}
\begin{eqnarray}
C_{k_2}=\frac{\sum_{i=1}^{p-1}{\Delta \overline{HPV_{i}(k)}}}{p-1}
\label{Delta HPV4}
\end{eqnarray}

$C_{k_1}$ and $C_{k_2}$ are the two criteria that are respectively used for measuring the sensitivity of $X(k)$ to the variations of relative state. A bigger value of $C_{k_1}$ or $C_{k_2}$ means a bigger sensitivity. We have sorted the HPVs measured by the pressure sensors from biggest to smallest according to $C_{k_1}$ and $C_{k_2}$, respectively. Then we have successively changed the number of pressure sensors used, that is, the number of variables ($M$) for establishing the regression model and thus ensuring that the pressure sensors used are not redundant or insufficient. For example, when $M=3$, we used the first $M$ pressure sensors for regression analysis.
\subsection{Importance Measurement of the HPVs Measured by Each Pressure Sensor}
Importance measurement is conducted for evaluating the importance of the HPV measured by a pressure sensor (variable $X(k) (k=1, \dots, p)$ in the feature vector $X$) to the relative state in RF-based analysis. The detailed processes of conducting importance measurement are as follows.

$\textbf{Step 1:}$ Obtain an initial random forest model, including $N$ trees. Then respectively calculate the mean-square error $(MSE)$ of the estimated results using $N$ OOB samples. And the obtained $MSE$s can be defined as $MSE_i (i=1, \dots, N)$, taking the form as
\begin{eqnarray}
MSE_i=\frac{\sum_{j=1}^{n}\left ( \hat{Y}_{i}(j)-Y_{i}(j) \right )^2}{n}
\end{eqnarray}
where $\hat{Y}_{i}(j)$ and $Y_{i}(j)$ indicate the $j-th$ estimated value and actual relative state value in the $i-th$ OOB sample, respectively.

$\textbf{Step 2:}$ Randomly permute $X(k) (k=1, \dots, p)$ of the feature vector $X$ in the OOB samples, for obtaining new OOB samples. Then calculate the $MSE_i(k) (i=1, \dots, N; k=1, \dots, p)$ of the estimated results using the new OOB samples. Basing on the above processes, a MSE matrix can be obtained, taking the form as
\begin{eqnarray}
\begin{bmatrix}
MSE_1(1) &MSE_2(1)  &\cdots   &MSE_N(1) \\
MSE_1(2) &MSE_2(2)  &\cdots   &MSE_N(2) \\
\vdots  &\vdots   &\vdots   & \\
MSE_1(p) &MSE_2(p)  &\cdots   &MSE_N(p) \\
\end{bmatrix}
\label{MSE matrix}
\end{eqnarray}

$\textbf{Step 3:}$ Calculate the difference between elements in $[MSE_1, \dots, MSE_N ]^T$ and $[MSE_i(1), \dots, MSE_i(p) ]^T$, and the difference can be defined as
\begin{eqnarray}
\Delta MSE_i(k)=MSE_i-MSE_i(k)
\end{eqnarray}
where $i=1, \dots, N$, and $k=1, \dots, p$.

$\textbf{Step 4:}$ Calculate the importance of variable $X(k) (k=1, \dots, p)$. The importance value $I_k$ is defined as
\begin{eqnarray}
I_k=\frac{\sum_{i=1}^{N}\Delta MSE_i(k)}{N\cdot SE_k}
\end{eqnarray}
where $SE_k$ is the standard error of $\Delta MSE_i(k) (i=1, \dots, N)$, taking the form as
 \begin{eqnarray}
SE_k=\sqrt{\frac{\sum_{i=1}^{N}(\Delta MSE_i(k)-\overline{MSE_k})^2}{N}}
\end{eqnarray}
where $\overline{MSE_k}$ is the mean of $\Delta MSE_i(k) (i=1, \dots, N)$.
A bigger $I_k$ indicates that $X(k)$ is more important in the regression model.
\subsection{Evaluation of the Regression Model}
Mean absolute error ($MAE$) and coefficient of determination ($R^2$) are used for evaluating the accuracy of the above-trained regression model. Smaller $MAE$ and bigger $R^2$ indicate a more accurate model.
\begin{eqnarray}
MAE=\frac{\sum_{j=1}^{n}\left | \hat{y_j}-y_{j} \right |}{n}
\label{MAE}
\end{eqnarray}
\begin{eqnarray}
R^2=1-\frac{\sum_{j=1}^{n}\left ( y_{j}-\hat{y_j} \right )^2}{\sum_{j=1}^{n}\left ( y_{j}-\bar{y} \right )^2}
\label{R2}
\end{eqnarray}
where $\hat{y_i}$ and $y_i$ indicates the estimated value and actual relative state value in the $i-th$ sample. $\bar{y}$ indicates the mean of $y_{j} (j=1, \dots, n)$.
\section{Results}
\subsection{Analytic Results of Insufficiency and Redundancy of the Pressure Sensors}
\begin{table}[]
\centering
\caption{$C_{k_1}$ of each pressure sensor in each experiment.}
\label{$C_{k_1}$ of each pressure sensor in each experiments}
\begin{tabular}{@{}cccccccc@{}}
\toprule
            & $\revision{d}$ & $A$ & $f$ & $\phi$ & $\alpha$ & $\beta$ & $\gamma$ \\ \midrule
$P_0$    &0.3165                         &0.0667   &0.2000   &0.1633                     &0.1691                       &0.2446                      &0.4535                      \\
$P_{L_1}$ &0.3117                         &0.1109   &0.2000   &0.1340                     &0.1268                       &0.2586                      &0.1495                       \\
$P_{L_2}$ &0.1672                         &0.2608   &0.4971   &0.2757                     &0.1420                       &0.1713                      &0.1205                       \\
$P_{L_3}$ &0.2676                         &0.3224   &0.3555   &0.2394                     &0.1336                       &0.2439                      &0.1000                       \\
$P_{L_4}$ &0.4260                         &0.3172   &0.3354   &0.2143                     &0.1113                       &0.3339                      &0.2280                       \\
$P_{R_1}$ &0.3280                         &0.0968   &0.2025   &0.1427                     &0.1205                       &0.2199                      &0.1708                       \\
$P_{R_2}$ &0.1723                         &0.2442   &0.5251   &0.2713                     &0.1276                       &0.1606                      &0.1167                       \\
$P_{R_3}$ &0.3130                         &0.3174   &0.4818   &0.2009                     &0.1346                       &0.3175                      &0.1000                       \\
$P_{R_4}$ &0.2750                         &0.3012   &0.5028   &0.2162                     &0.1030                       &0.2462                      &0.1009                       \\\bottomrule
\end{tabular}
\end{table}
\begin{table}[]
\centering
\caption{Sorting the order of the HPVs measured by the pressure sensors according to $C_{k_1}$ (from biggest to smallest).}
\label{Sorting the order of the HPVs measured by the pressure sensors according to $C_{k_1}$}
\begin{tabular}{@{}cccccccc@{}}
\toprule
Experiments &Order  \\ \midrule
$\revision{d}$ &$P_{L_4}$, $P_{R_1}$, $P_{0}$, $P_{R_3}$, $P_{L_1}$, $P_{R_4}$, $P_{L_3}$, $P_{R_2}$, $P_{L_2}$  \\
$A$ &$P_{L_3}$, $P_{R_3}$, $P_{L_4}$, $P_{R_4}$, $P_{L_2}$, $P_{R_2}$, $P_{L_1}$, $P_{R_1}$, $P_{0}$    \\
$f$ &$P_{R_2}$, $P_{R_4}$, $P_{L_2}$, $P_{R_3}$, $P_{L_3}$, $P_{L_4}$, $P_{R_1}$, $P_{0}$, $P_{L_1}$   \\
$\phi$ &$P_{L_2}$, $P_{R_2}$, $P_{L_3}$, $P_{R_4}$, $P_{L_4}$, $P_{R_3}$, $P_{0}$, $P_{R_1}$, $P_{L_1}$\\
$\alpha$ &$P_{0}$, $P_{L_2}$, $P_{R_3}$, $P_{L_3}$, $P_{R_2}$, $P_{L_1}$, $P_{R_1}$, $P_{L_4}$, $P_{R_4}$   \\
$\beta$ &$P_{L_4}$, $P_{R_3}$, $P_{L_1}$, $P_{R_4}$, $P_{0}$, $P_{L_3}$, $P_{R_1}$, $P_{L_2}$, $P_{R_2}$  \\
$\gamma$ &$P_{0}$, $P_{L_4}$, $P_{R_1}$, $P_{L_1}$, $P_{L_2}$, $P_{R_2}$, $P_{R_4}$, $P_{L_3}$, $P_{R_3}$  \\ \bottomrule
\end{tabular}
\end{table}
\begin{table}[]
\centering
\caption{$C_{k_2}$ of each pressure sensor in each experiment.}
\label{$C_{k_2}$ of each pressure sensor in each experiments}
\begin{tabular}{@{}cccccccc@{}}
\toprule
            & $\revision{d}$ & $A$ & $f$ & $\phi$ & $\alpha$ & $\beta$ & $\gamma$ \\ \midrule
$P_0$    &11.3143                         &3.4602   &2.5864   &3.8769                     &5.2647                       &6.9166                      &0.9537                      \\
$P_{L_1}$ &2.3996                         &1.6968   &1.7521   &2.4489                     &4.4564                       &3.0227                      &1.6151                       \\
$P_{L_2}$ &1.4778                         &1.3393   &1.1334   &1.1984                     &2.0292                       &1.6177                      &1.0072                       \\
$P_{L_3}$ &0.6341                         &1.2951   &0.8205   &1.3811                     &1.1658                       &1.6542                      &0.5840                       \\
$P_{L_4}$ &1.0560                         &1.4444   &0.4281   &1.1442                     &1.2445                       &1.6735                      &0.2736                       \\
$P_{R_1}$ &2.2620                         &1.4399   &1.5111   &2.1011                     &4.1315                       &2.9065                      &1.4136                       \\
$P_{R_2}$ &1.7180                         &1.2739   &0.8769   &1.1659                     &2.0737                       &1.9250                      &0.9149                       \\
$P_{R_3}$ &0.7541                         &1.3676   &0.6749   &1.1011                     &1.2411                       &1.0540                      &0.4223                       \\
$P_{R_4}$ &0.6398                         &1.2407   &0.9156   &0.9262                     &1.2121                       &1.5360                      &0.1482                       \\\bottomrule
\end{tabular}
\end{table}
\begin{table}[]
\centering
\caption{Sorting the order of the HPVs measured by the pressure sensors according to $C_{k_2}$ (from biggest to smallest)}
\label{Sorting the order of the HPVs measured by the pressure sensors according to $C_{k_2}$}
\begin{tabular}{@{}cccccccc@{}}
\toprule
Experiments &Order  \\ \midrule
$\revision{d}$ &$P_{0}$, $P_{L_1}$, $P_{R_1}$, $P_{R_2}$, $P_{L_2}$, $P_{L_4}$, $P_{R_3}$, $P_{R_4}$, $P_{L_3}$  \\
$A$ &$P_{0}$, $P_{L_1}$, $P_{L_4}$, $P_{R_1}$, $P_{R_3}$, $P_{L_2}$, $P_{L_3}$, $P_{R_2}$, $P_{R_4}$    \\
$f$ &$P_{0}$, $P_{L_1}$, $P_{R_1}$, $P_{L_2}$, $P_{R_4}$, $P_{R_2}$, $P_{L_3}$, $P_{R_3}$, $P_{L_4}$   \\
$\phi$ &$P_{0}$, $P_{L_1}$, $P_{R_1}$, $P_{L_3}$, $P_{L_2}$, $P_{R_2}$, $P_{L_4}$, $P_{R_3}$, $P_{R_4}$\\
$\alpha$ &$P_{0}$, $P_{L_1}$, $P_{R_1}$, $P_{R_2}$, $P_{L_2}$, $P_{L_4}$, $P_{R_3}$, $P_{R_4}$, $P_{L_3}$   \\
$\beta$ &$P_{0}$, $P_{L_1}$, $P_{R_1}$, $P_{R_2}$, $P_{L_4}$, $P_{L_3}$, $P_{L_2}$, $P_{R_4}$, $P_{R_3}$  \\
$\gamma$ &$P_{L_1}$, $P_{R_1}$, $P_{L_2}$, $P_{0}$, $P_{R_2}$, $P_{L_3}$, $P_{R_3}$, $P_{L_4}$, $P_{R_4}$  \\ \bottomrule
\end{tabular}
\end{table}
\begin{figure}[!h]
\centering
\includegraphics[width=\linewidth]{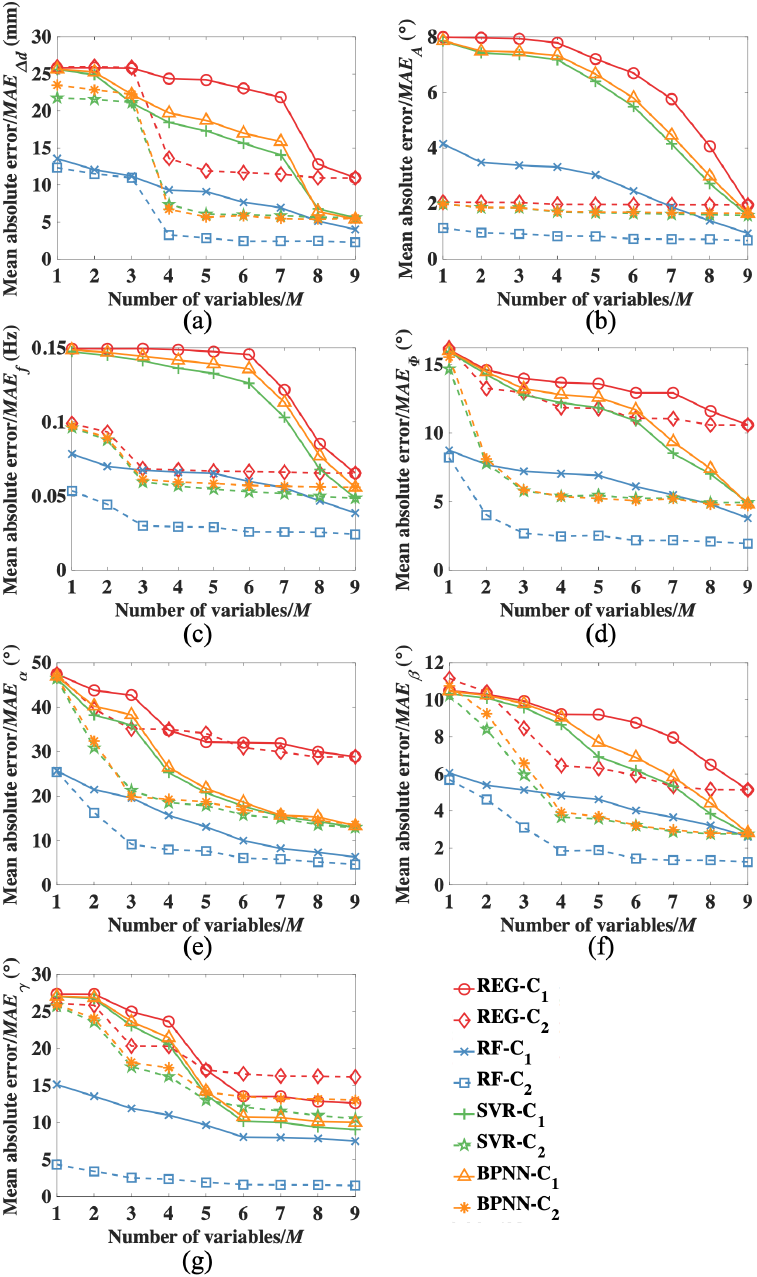}
\caption{\revision{Mean absolute error ($MAE$) with the varied number of variables $M$ when using RF algorithm, BPNN, SVR method, and REG method. Take REG-C$_1$ for example, it refers to the result obtained by the REG method based on the order of the pressure sensor sorted by $C_{k_1}$.}}
\label{Mean absolute error}
\end{figure}
\begin{figure}[!h]
\centering
\includegraphics[width=\linewidth]{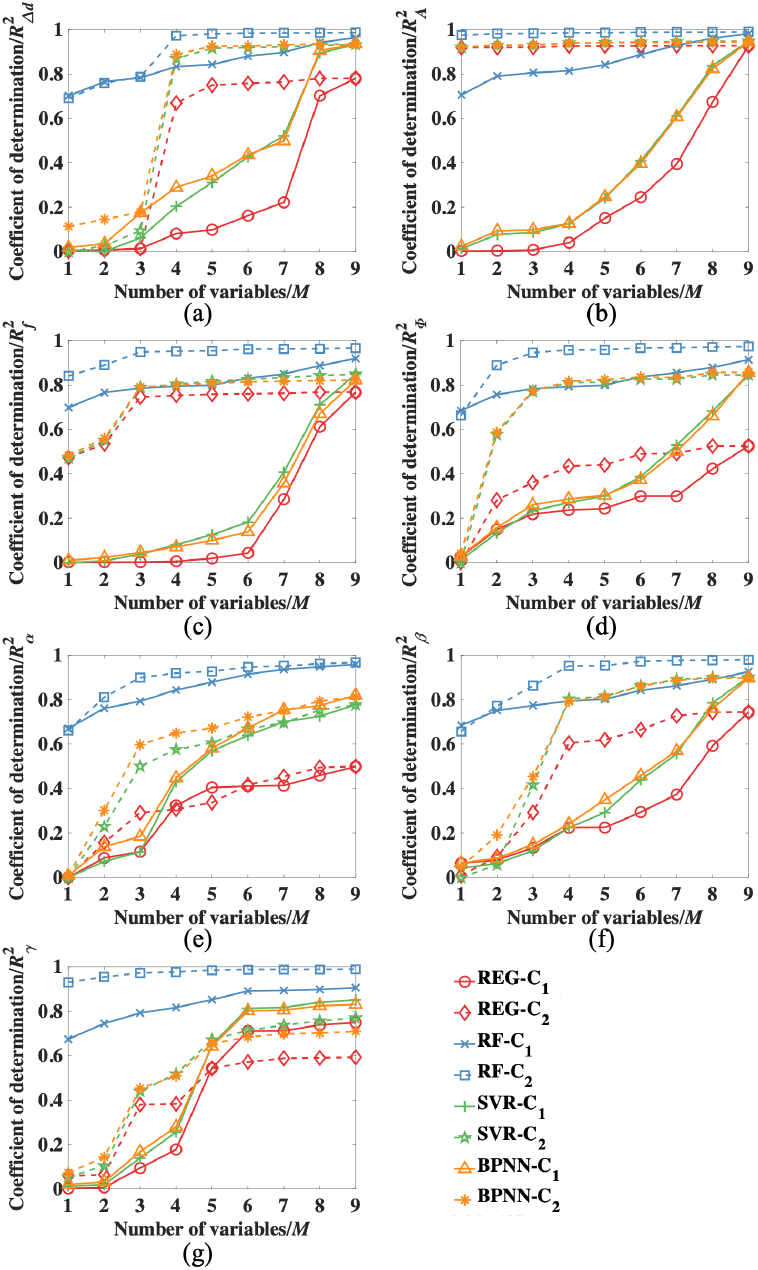}
\caption{\revision{Coefficient of determination ($R^2$) with the varied number of variables $M$ when using RF algorithm, BPNN, SVR method, and REG method.}}
\label{Coefficient of determination}
\end{figure}
Table~\ref{$C_{k_1}$ of each pressure sensor in each experiments} and Table~\ref{$C_{k_2}$ of each pressure sensor in each experiments} show $C_{k_1}$ and $C_{k_2}$ of each pressure sensor in each experiments. Table~\ref{Sorting the order of the HPVs measured by the pressure sensors according to $C_{k_1}$} and Table~\ref{Sorting the order of the HPVs measured by the pressure sensors according to $C_{k_2}$} show the order of the HPVs measured by the pressure sensors from biggest to smallest based on the value of $C_{k_1}$ and $C_{k_2}$. It can be seen that there are significant differences between the two orders of the HPVs. In the experiments, fish head of the downstream robot directly faces to the vortex wake generated by the upstream oscillating caudal fin, so the vortex wake has a more significant effect on the fish head. And thus, the HPVs measured by the pressure sensors around the fish head are significantly bigger than HPVs measured by pressure sensors at the posterior. In order to ensure the comparability of different pressure sensors, we have nondimensionalized the $\Delta \overline{HPV_{i}(k)}$ and then calculate $C_{k_1}$. On the other hand, we have also calculated calculate $C_{k_2}$ using the $\Delta \overline{HPV_{i}(k)}$, which has not been nondimensionalized. The comparisons of regression results corresponding to $C_{k_1}$ and $C_{k_2}$ have been conducted in the following part.
\subsection{Regression Results Using the Four Methods}
\begin{figure*}[!h]
\centering
\includegraphics[width=\linewidth]{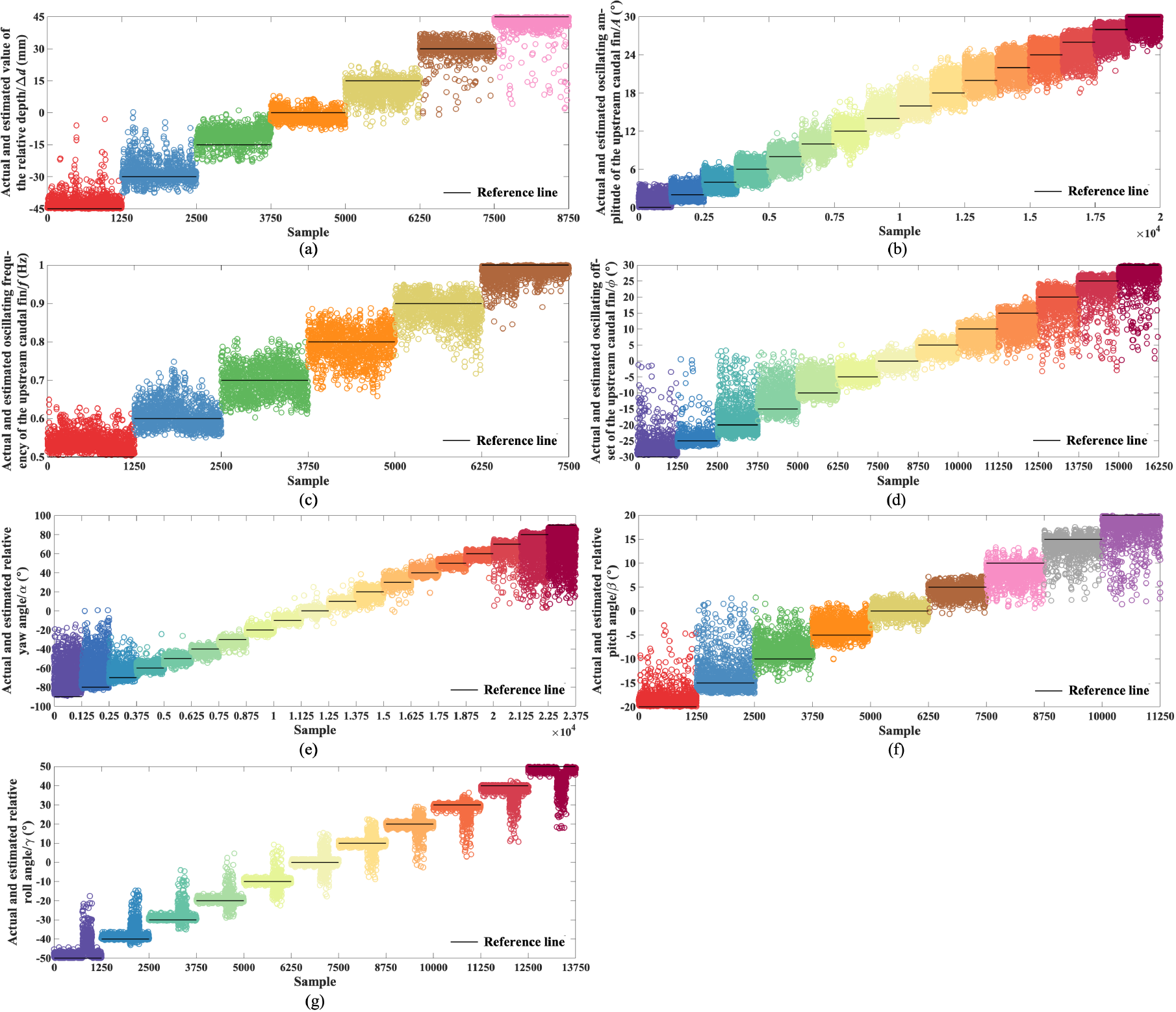}
\caption{\revision{The actual relative states and estimated values obtained by the RF algorithm based regression model.}}
\label{The estimated and actual relative states obtained by random forest method}
\end{figure*}
Figure~\ref{Mean absolute error} shows the mean absolute error $MAE$ with the varied number of variables $M$, that is, the number of pressure sensors used. It can be seen that $MAE$ decreases with the increasing $M$ as a whole. Figure~\ref{Coefficient of determination} shows the coefficient of determination $R^2$ with the varied number of variables $M$. $R^2_{d}$, $R^2_A$, $R^2_f$, $R^2_A$, $R^2_\phi$, $R^2_\alpha$, $R^2_\beta$, and $R^2_\gamma$ indicates $R^2$ in the experiments of investigating $\revision{d}$, $A$, $f$, $\phi$, $\alpha$, $\beta$, and $\gamma$. It can be seen that $R^2$ increases with the number of $M$ as a whole in each experiment when using the four methods. Comparing the $MAE$ and $R^2$ obtained based on the order of HPVs sorted by $C_{k_1}$ and $C_{k_2}$, it can be seen that regression models obtained according to order of HPVs sorted by $C_{k_2}$ have better effects. From this perspective, $C_{k_2}$ is more reasonable for measuring the sensitivity of the HPVs measured by pressure sensors to the variations of relative states. \revision{In the following part, we mainly focus on the results corresponding to $C_{k_2}$.}

Take the experiment of investigating $\revision{d}$ for example, a more careful inspection reveals that $R^2$ and $MAE$ vary little when $M$ exceeds 4.  This characteristic demonstrates that 4 pressure sensors which specifically refer to $P_{0}$, $P_{L_1}$, $P_{R_1}$, and $P_{R_2}$ here are enough for regression analysis. In other words, five or more pressure sensors are redundant. In this way, the reasonable numbers (defined as $M_r$) of pressure sensors used for regression analyses of $\revision{d}$, $A$, $f$, $\phi$, $\alpha$, $\beta$, and $\gamma$ are determined as 4, 1, 3, 4, 7, 4, and 5, respectively.

On the other hand, in each experiment, $R^2$s obtained by the RF algorithm is bigger than $R^2$s obtained by BPNN, REG, and SVR. A further inspection has revealed that $R^2$ obtained by RF has the best performance though $M$ is small. Basing on the above analyses, we can conclude that the RF algorithm has better regression effects than the other three methods. Besides, $R^2$ obtained by the RF algorithm varies smoothly while the $R^2$ obtained by the other three methods have more significant variations with $M$. Such a characteristic demonstrates that random forest has better noise-resistibility.

On the above analyses, a conclusion can be obtained that RF algorithm has the best regression effect. Figure~\ref{The estimated and actual relative states obtained by random forest method} shows the estimated and actual relative states obtained by the RF algorithm with $M_r=$4, 1, 3, 4, 7, 4, and 5 for the experiment of $\revision{d}$, $A$, $f$, $\phi$, $\alpha$, $\beta$, and $\gamma$, respectively. On the whole, it can be seen that the established RF based regression models have good performances for describing the relative states. Specifically, the combination of the $R^2$, $MAE$, and $M_r$ corresponding to the best regression effect is defined as ($R^2$, $MAE$, $M$), which refers to (0.972, 3.250 mm, 4), (0.975, 1.119$^\circ$, 1), (0.949, 0.030 Hz, 3), (0.958, 2.467$^\circ$, 4), (0.952, 5.778$^\circ$, 7), (0.952, 1.836$^\circ$, 4), and (0.985, 1.915$^\circ$, 5) for the experiment of $\revision{d}$, $A$, $f$, $\phi$, $\alpha$, $\beta$, and $\gamma$, respectively.

\begin{figure}[!h]
\centering
\includegraphics[width=0.85\linewidth]{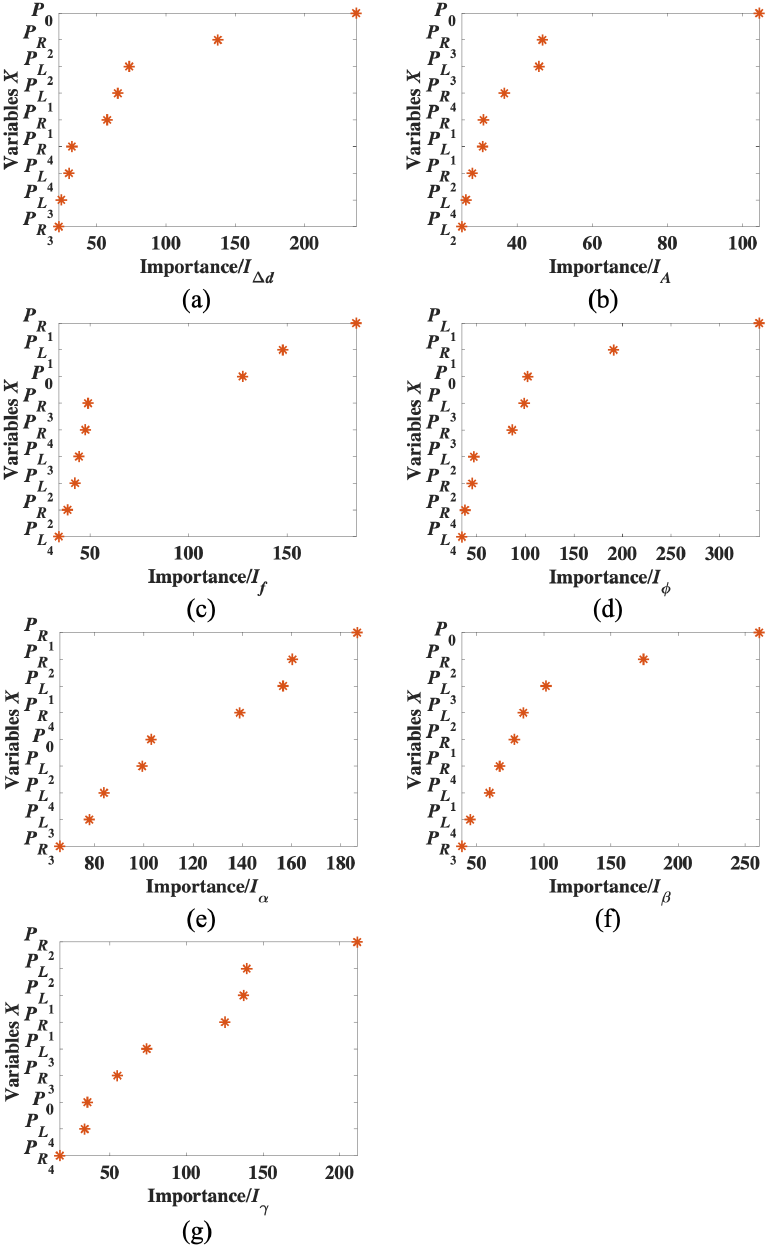}
\caption{\revision{Variables $X(k)$ order sorted by the importance $I_k$.}}
\label{Importance}
\end{figure}
Figure~\ref{Importance} shows the order of the variables $X(k)$ sorted by the importance $I_k$. A bigger $I_k$ indicates that the variable $X(k)$ is more important to the relative state in the RF-based regression analysis. The combinations of the $M_r$ most important variables for $\revision{d}$, $A$, $f$, $\phi$, $\alpha$, $\beta$, and $\gamma$ are ($P_{0}$, $P_{R_2}$, $P_{L_2}$, $P_{L_1}$), ($P_0$), ($P_{R_1}$, $P_{L_1}$, $P_{0}$), ($P_{L_1}$, $P_{R_1}$, $P_{0}$, $P_{L_3}$ ), ($P_{R_1}$, $P_{R_2}$, $P_{L_1}$, $P_{R_4}$, $P_{0}$, $P_{L_2}$, $P_{L_4}$), ($P_{0}$, $P_{R_2}$, $P_{L_3}$, $P_{L_2}$), and ($P_{R_2}$, $P_{L_2}$, $P_{L_1}$, $P_{R_1}$, $P_{L_3}$), respectively. A further investigation has revealed that $M_r$ pressure sensors in the above combinations are almost the first $M_r$ pressure sensors in Table~\ref{Sorting the order of the HPVs measured by the pressure sensors according to $C_{k_2}$} for each experiment. Such a characteristic demonstrates also proves that it is reasonable for sorting the pressure sensors according to $C_{k_2}$ because the  first $M_r$ pressure sensors in Table~\ref{Sorting the order of the HPVs measured by the pressure sensors according to $C_{k_2}$} happen to be \revision{the} most important in the regression analyses. Besides, there exist differences between concrete orders of pressure sensors in the above combinations and Table~\ref{Sorting the order of the HPVs measured by the pressure sensors according to $C_{k_2}$}. Such a characteristic demonstrates that the sensitivity of the pressure sensors-measured HPVs to the relative states is not the same thing as the importance of the HPVs.
\subsection{Random Forest Algorithm Based Relative Yaw Angle Estimation and Oscillating Amplitude Estimation}
\begin{figure}[!h]
\centering
\includegraphics[width=\columnwidth]{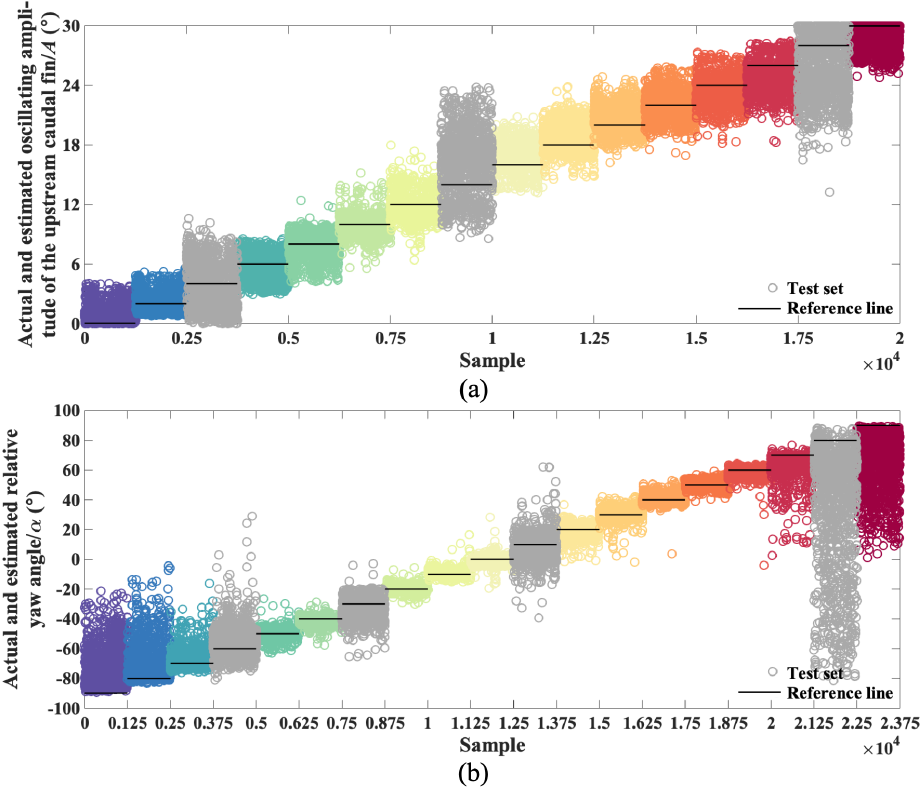}
\caption{\revision{Results of RF-based relative yaw angle and oscillating amplitude estimation. (a) Oscillating amplitude estimation ($M_r=$1, the pressure sensor used is $P_0$). (b) Relative yaw angle estimation ($M_r=$7, the pressure sensors used are $P_{0}$, $P_{L_1}$, $P_{R_1}$, $P_{R_2}$, $P_{L_2}$, $P_{L_4}$, and $P_{R_3}$).}}
\label{Results of RF based relative yaw angle and oscillating amplitude estimation}
\end{figure}

The above work has shown that RF method has the best regression effect. In this part, we validate the effectiveness of RF method in relative state estimation. Considering that the number of the investigated experimental parameters, $\revision{d}$, $A$, $f$, $\phi$, $\alpha$, $\beta$, and $\gamma$, is 7, 16, 6, 13, 19, 9, and 11, respectively. Here, we select the experiments of investigating $A$ and $\alpha$ for validation works because more experimental parameters have been considered. Specifically, 80\% data in the data set $O$ form a training set for training the RF-based models, and the remaining 20\% data form a test set, for validating the effect of the RF-based models in estimating the relative yaw angle and the oscillating amplitude. Figure~\ref{Results of RF based relative yaw angle and oscillating amplitude estimation} shows the results of RF-based oscillating amplitude and relative yaw angle estimation.
$R^2$ and $MAE$ of the regression model for oscillating amplitude estimation is 0.975 and 1.091$^\circ$, respectively. And the $MAE$s of the oscillating amplitude obtained by test set are 2.171$^\circ$, 3.163$^\circ$, and 3.211$^\circ$ for $A=4^\circ, 14^\circ, 28^\circ$, respectively. $R^2$ and $MAE$ of the regression model for the relative yaw angle is 0.956 and 5.513$^\circ$, respectively. And the $MAE$s of the relative yaw angle obtained by test set are 9.284$^\circ$, 7.506$^\circ$, 7.694$^\circ$, and 45.435$^\circ$ for $\alpha=-60^\circ, -30^\circ, 30^\circ, 60^\circ$, respectively. It can be seen that the regression model has a bad performance when the relative yaw angle is big enough. However, it performs well when the relative yaw angle is small. The estimation errors may mainly result from the low noise-signal ratio of the HPVs measured by the pressure sensors. Though the pretreatment of the HPVs has been conducted, the hardware deficiency of the pressure sensors has reduced the estimation accuracy. In this case, improving the ALLS is necessary and urgent for improving the data quality. Besides, further data pretreatment by fusing ALLS data and IMU data may also provide a potential way for improving the estimation accuracy. The above-mentioned will be conducted in the following researches on multiple adjacent freely-swimming robotic fish.
\section{Conclusions and Future Work}
This article has investigated the regression model, which links the ALLS-measured HPVs to the relative states between two adjacent robotic fish with leader-follower formation in a flume. \revision{Two criteria are proposed firstly for investigating not only the sensitivity of each pressure sensor to the variations of relative state variations, but also the insufficiency and redundancy of the pressure sensors used for regression analysis.} Then four methods, including RF algorithm, BPNN, SVR, and REG, are used for establishing the regression model. Comparisons of the effects of the regression effects of using the four methods have been conducted, for determining the best method. The results show that the RF algorithm has the best excellent performance in estimating the relative states using the ALLS-measured HPVs. And RF-based regression models have been established for predicting the relative yaw angle and the oscillating amplitude of the upstream robotic fish with small errors. Moreover, a further discussion for the flume experiments has been conducted in detail.

\revision{For different underwater vehicles like propeller-actuated autonomous underwater vehicle (AUV) and remotely operated vehicle (ROV), the flow variations caused by the propeller and vehicle motions are essentially different from fin-actuated robot. In this article, we have selected a specifically designed mechatronic system (robotic fish) for investigating artificial lateral line system based relative state estimation for underwater robots group. The regression methods used in this article could be used in other underwater mechatronic systems.}

In the future, we will conduct an online estimation of the relative states between two adjacent freely-swimming robotic fish using the RF-based regression model. Free motions of the robotic fish result in rhythmical oscillations of the fish body, which may have significant effects on the ALLS-measured HPVs \cite{zheng2019artificial}. So the qualitative and quantitative relationships between the HPVs and the relative states may significantly differ from those we have investigated in this article and \cite{Wang2015Sensing,zheng2017artificial}. Relative state estimation for two adjacent freely-swimming robotic fish would be more challenging.
\section*{Acknowledgment}
This work was supported in part by grants from the National Natural Science Foundation of China (NSFC, No. 91648120, 61633002, 51575005) and the Beijing Natural Science Foundation (No. 4192026).
\ifCLASSOPTIONcaptionsoff
  \newpage
\fi

\bibliographystyle{IEEEtran}
\bibliography{lateral_line}

\begin{IEEEbiography}[{\includegraphics[width=1in,height=1.25in,clip,keepaspectratio]{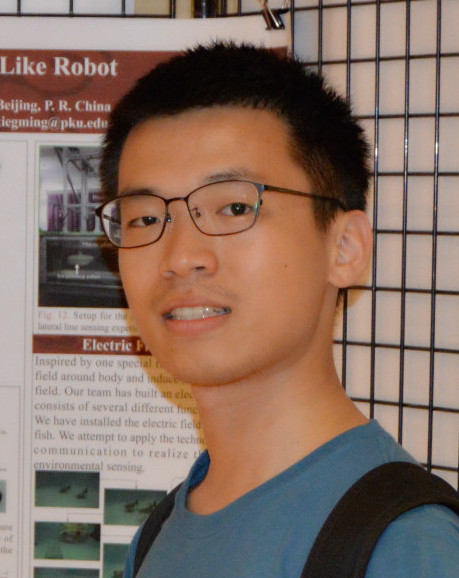}}]{Xingwen Zheng}
received the B.E. in Mechanical Engineering and Automation from Northeastern University, Shenyang, China in 2015. He is currently a PhD candidate at the Intelligent Biomimetic Design Lab, State Key Laboratory for Turbulence and Complex Systems, College of Engineering, Peking University, Beijing, China. His current research interests include biomimetic robotics, lateral line inspired sensing, and multi-robot control.
\end{IEEEbiography}
\begin{IEEEbiography}[{\includegraphics[width=1in,height=1.25in,clip,keepaspectratio]{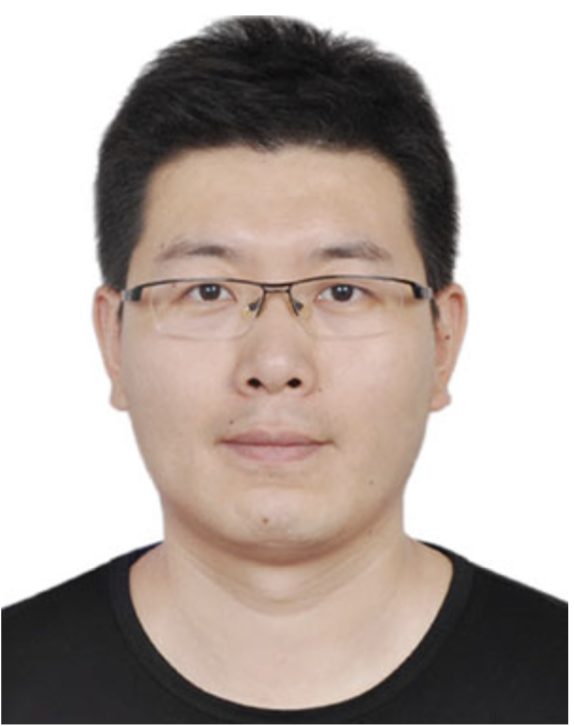}}]{Wei Wang}
received the B.E. degree in automation from the University of Electronic Science and Technology of China, Chengdu, Sichuan, China, and the Ph.D. degree in general mechanics and foundation of mechanics from Peking University, Beijing, China, in 2010 and 2016, respectively. He is currently a joint Postdoctoral Researcher with the Senseable City Laboratory and the Computer Science and Artificial Intelligence Laboratory, Massachusetts Institute of
Technology, Cambridge, MA, USA. His current research interests include bio-inspired robots, driverless vehicles, and collective intelligence.
\end{IEEEbiography}
\begin{IEEEbiography}[{\includegraphics[width=1in,height=1.25in,clip,keepaspectratio]{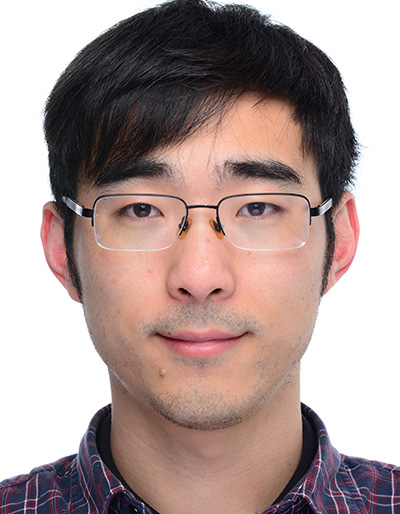}}]{Liang Li}
received the B. E. degree in automation from Chongqing University, Chongqing, China, and the Ph. D. degree in General Mechanics and Foundation of Mechanics from Peking University, Beijing China, in 2011 and 2017, respectively. He is currently a Post-Doctoral Research Fellow in the Department of Collective Behaviour, Max-Planck Institute for Ornithology, Konstanz, Germany. His current research interests include collective behaviour in hybrid animal-robot systems, biofluid dynamics in fish school and swarm intelligence in robots.
\end{IEEEbiography}
\begin{IEEEbiography}[{\includegraphics[width=1in,height=1.25in,clip,keepaspectratio]{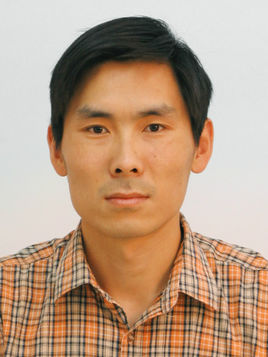}}]{Guangming Xie}
received his B.S. degrees in both Applied Mathematics and Electronic and Computer Technology, his M.E. degree in Control Theory and Control Engineering, and his Ph.D. degree in Control Theory and Control Engineering from Tsinghua University, Beijing, China in 1996, 1998, and 2001, respectively. Then he worked as a postdoctoral research fellow in the Center for Systems and Control, Department of Mechanics and Engineering Science, Peking University, Beijing, China from July 2001 to June 2003. In July 2003, he joined the Center as a lecturer. Now he is a Full Professor of Dynamics and Control in the College of Engineering, Peking University.

He is an Associate Editor of Scientific Reports, International Journal of Advanced Robotic Systems, Mathematical Problems in Engineering and an Editorial Board Member of Journal of Information and Systems Science. His research interests include smart swarm theory, multi-agent systems, multi-robot cooperation, biomimetic robot, switched and hybrid systems, and networked control systems.
\end{IEEEbiography}
\end{document}